%% file: noniid-foltr.tex
\documentclass[sigconf]{acmart}
\pdfoutput=1

\usepackage{color}
\usepackage{subfigure}
\usepackage{amsmath}
\usepackage{tabularx}
\usepackage{algorithm}
\usepackage[noend]{algorithmic}
\usepackage{hyperref}
\newcommand{\SUB}[1]{\ENSURE \hspace{-0.15in} \textbf{#1}}

\newcommand{\dubbelop}{$^{\blacktriangle}$}

\usepackage{tikz}
\usepackage{xcolor}
\newcommand*\circled[1]{\tikz[baseline=(char.base)]{
		\node[shape=circle,fill,inner sep=0.6pt] (char) {\bf\small\textcolor{white}{#1}};}}

\AtBeginDocument{%
  \providecommand\BibTeX{{%
    \normalfont B\kern-0.5em{\scshape i\kern-0.25em b}\kern-0.8em\TeX}}}

\copyrightyear{2022} 
\acmYear{2022} 
\setcopyright{acmlicensed}\acmConference[SIGIR '22]{Proceedings of the 45th International ACM SIGIR Conference on Research and Development in Information Retrieval}{July 11--15, 2022}{Madrid, Spain}
\acmBooktitle{Proceedings of the 45th International ACM SIGIR Conference on Research and Development in Information Retrieval (SIGIR '22), July 11--15, 2022, Madrid, Spain}
\acmPrice{15.00}
\acmDOI{10.1145/3477495.3531709}
\acmISBN{978-1-4503-8732-3/22/07}

\begin{document}
\fancyhead{}


\title{Is Non-IID Data a Threat in Federated Online Learning to Rank?}

\author{Shuyi Wang}
\affiliation{
	\institution{The University of Queensland}
	\streetaddress{4072 St Lucia}
	\city{Brisbane}
	\state{QLD}
	\country{Australia}}
\email{shuyi.wang@uq.edu.au}

\author{Guido Zuccon}
\affiliation{
	\institution{The University of Queensland}
	\streetaddress{4072 St Lucia}
	\city{Brisbane}
	\state{QLD}
	\country{Australia}}
\email{g.zuccon@uq.edu.au}


\begin{abstract}
	
In this perspective paper we study the effect of non independent and identically distributed (non-IID) data on federated online learning to rank (FOLTR) and chart directions for future work in this new and largely unexplored research area of Information Retrieval. In the FOLTR process, clients participate in a federation to jointly create an effective ranker from the implicit click signal originating in each client, without the need to share data (documents, queries, clicks). A well-known factor that affects the performance of federated learning systems, and that poses serious challenges to these approaches, is that there may be some type of bias in the way data is distributed across clients. While FOLTR systems are on their own rights a type of federated learning system, the presence and effect of non-IID data in FOLTR has not been studied. To this aim, we first enumerate possible data distribution settings that may showcase data bias across clients and thus give rise to the non-IID problem. Then, we study the impact of each setting on the performance of the current state-of-the-art FOLTR approach, the Federated Pairwise Differentiable Gradient Descent (FPDGD), and we highlight which data distributions may pose a problem for FOLTR methods. We also explore how common approaches proposed in the federated learning literature address non-IID issues in FOLTR. This allows us to unveil new research gaps that, we argue, future research in FOLTR should consider. This is an important contribution to the current state of FOLTR field because, for FOLTR systems to be deployed, the factors affecting their performance, including the impact of non-IID data, need to be thoroughly understood.

\end{abstract}





\maketitle

\input{sections/introduction.tex}
\input{sections/related_works.tex}

\input{sections/foltr.tex}

\input{sections/noniid_data.tex}

\input{sections/conclusions.tex}

\begin{acks}
We would like to thank the anonymous reviewers for their insightful feedback in further shaping the paper. We would also like to thank Dr Bevan Koopman and Dr Harrisen Scells for their thoughtful comments on earlier drafts of this paper.
\end{acks}

\bibliographystyle{ACM-Reference-Format}
\bibliography{noniid-foltr}

%\appendix

\section*{Appendix} \label{sec-appendix}

Table~\ref{tbl:sdbn} reports the values of the parameters of the \emph{SDBN} click models we used in the experiments.

\begin{figure}[t]
	\centering
	\subfigure[\textbf{MSLR-WEB10k (linear ranker) under SDBN clicks}] { \label{fig:sdbn-linear} \includegraphics[width=7cm]{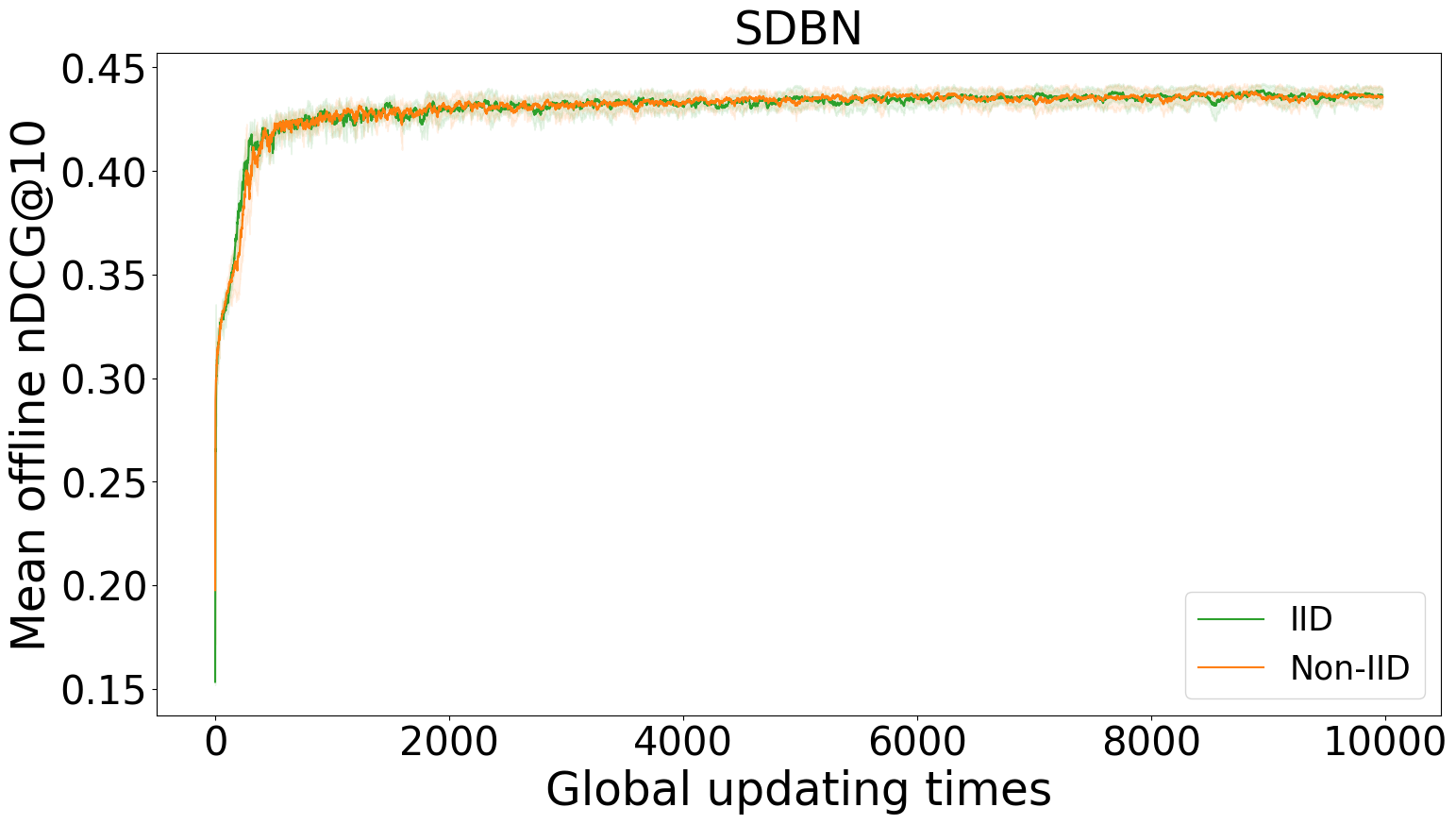}} 
	\subfigure[\textbf{MSLR-WEB10k (linear ranker) under PBM clicks}] { \label{fig:pbm-linear} \includegraphics[width=7cm]{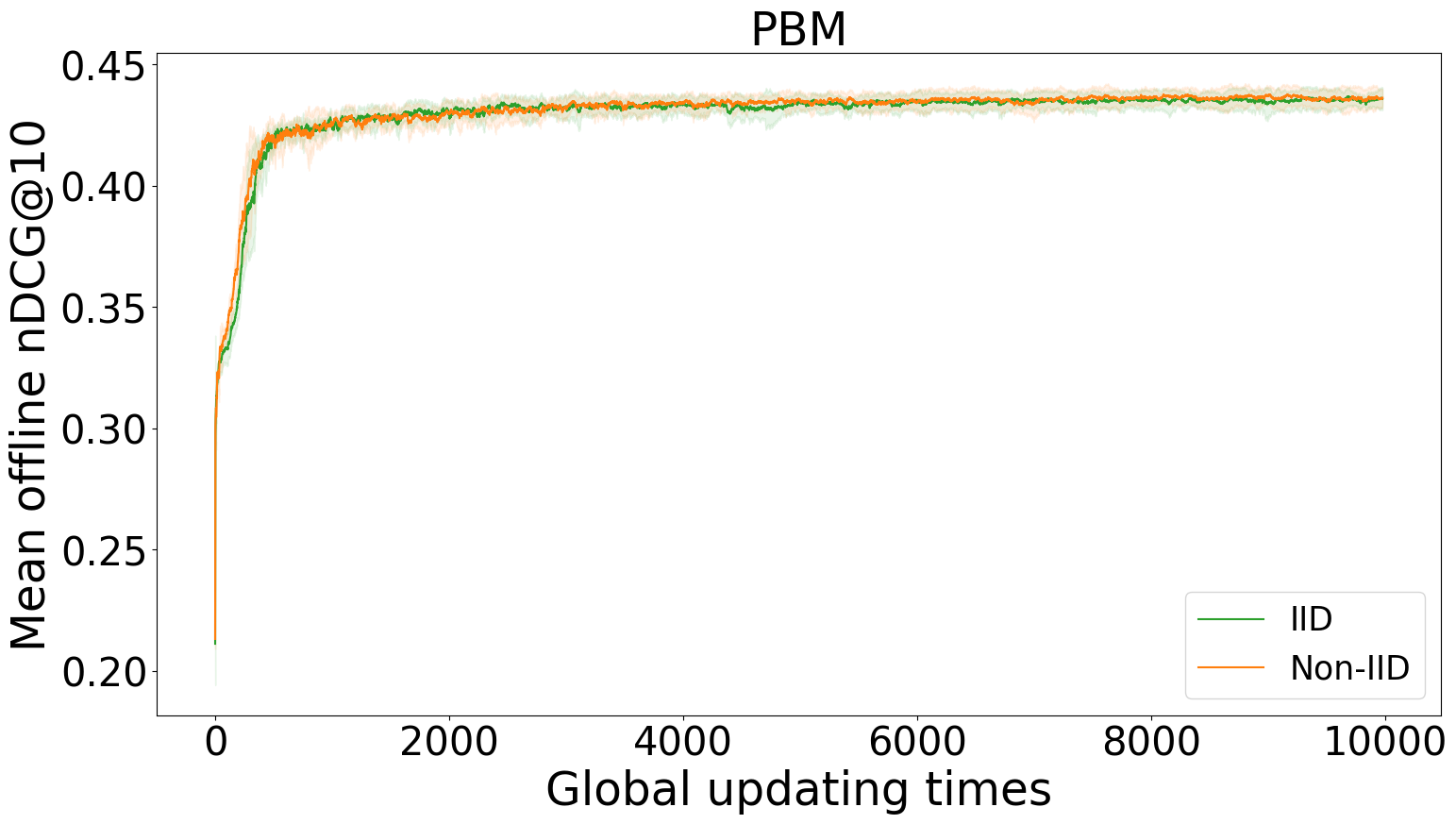}} 
	\vspace{-12pt}
	\caption{Offline performance (nDCG@10) on MSLR-WEB10k for Type 3, separately under SDBN and PBM click model; results averaged across all dataset splits and experimental runs.
	\label{fig:cps}}
\end{figure}

Figure~\ref{fig:cps} reports the results of our experiments for non-IID data type 3: click preferences.

\begin{figure*}[t]
	\centering
	\subfigure[\textbf{MSLR-WEB10k (linear ranker)}] {\label{fig:mslr-linear} \includegraphics[width=17cm]{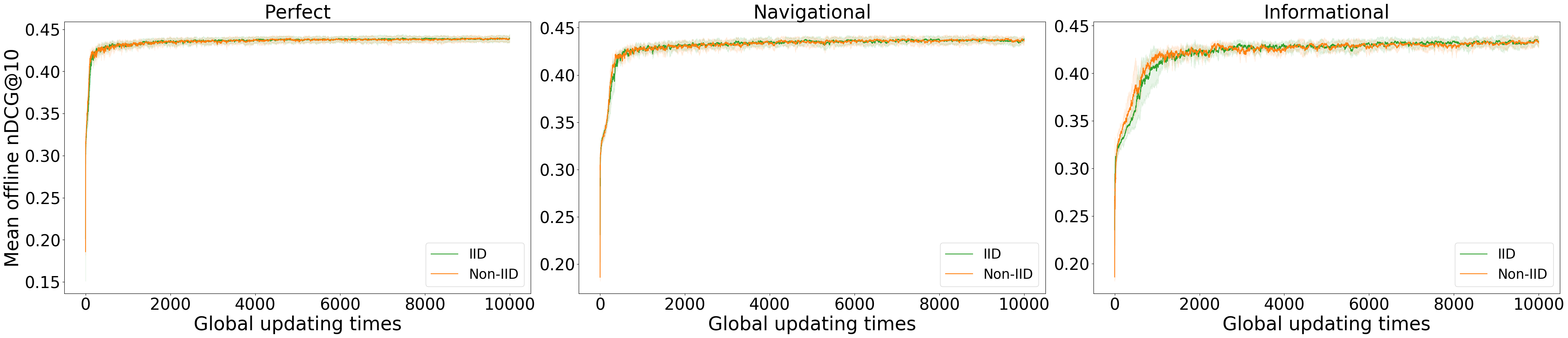}}
	\subfigure[\textbf{intent-change (linear ranker) mixed with type 1}] {\label{fig:intent-mixed-linear} \includegraphics[width=17cm]{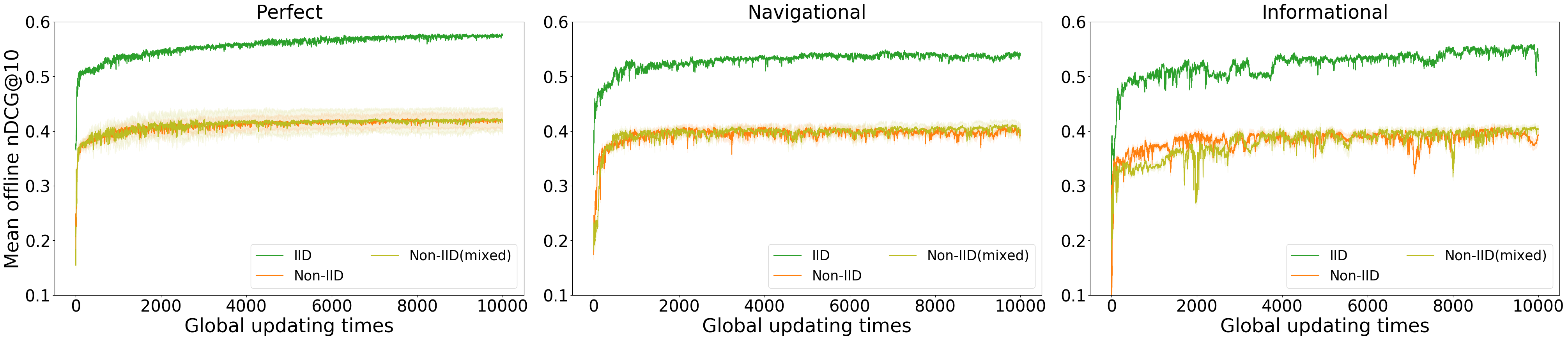}} 
	\subfigure[\textbf{MSLR-WEB10k (linear ranker) mixed with type 2 ($\# R = 1$)}] {\label{fig:mslr-mixed-linear}
	\includegraphics[width=17cm]{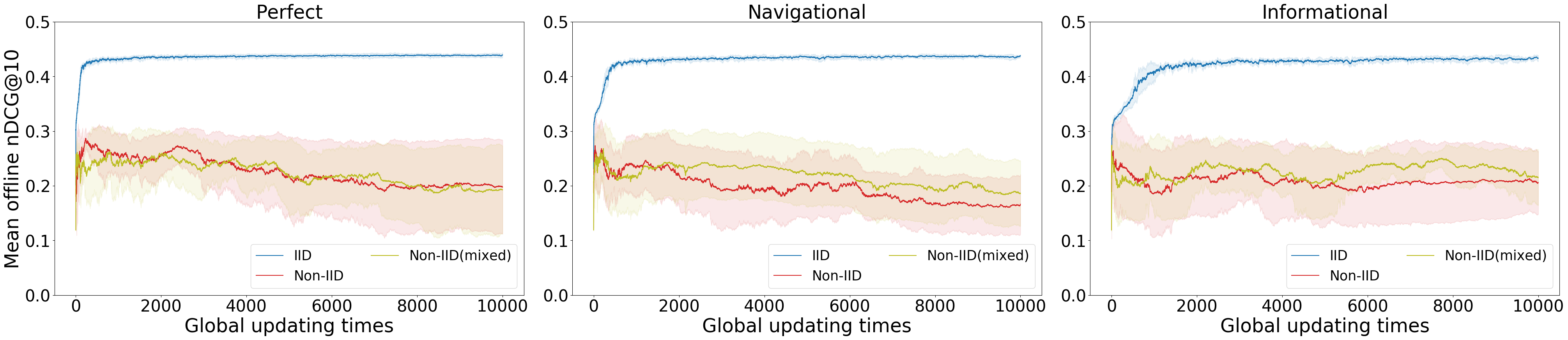}} 
	\vspace{-12pt}
	\caption{Offline performance (nDCG@10) on MSLR-WEB10k and intent-change for Type 4, under three instantiations of SDBN click model; results averaged across all dataset splits and experimental runs.
		\label{fig:dqs}}
\end{figure*}

Figure~\ref{fig:dqs} reports the results of our experiments for non-IID data type 4: data quantity.

For both type 3 and type 4 data experiments, as well as for other data types, the interested reader can find additional analysis and figures in the online appendix available at \url{https://github.com/ielab/non-iid-foltr}.

\begin{table}[t]
	\centering
	\caption[centre]{Instantiations of SDBN click model for simulating user behaviour in experiments. $rel(d)$ denotes the relevance label for document $d$. Note that in the intent-change dataset, only two-levels of relevance are used. We demonstrate the values for intent-change in bracket.}
	\begin{tabular}{ p{1.5cm}<{\centering} p{1.3cm}<{\centering} p{1.15cm}<{\centering} p{0.75cm}<{\centering} p{0.75cm}<{\centering} p{0.85cm}<{\centering}}
		\hline
		& \multicolumn{5}{c}{$P(\mathit{click}=1\mid rel(d))$}  \\
		%\hline
		\cmidrule(r){2-6}
		\emph{rel(d)} & 0 & 1 & 2 & 3 & 4 \\
		\hline
		\emph{perfect} &  0.0 (0.0)&  0.2 (1.0)&  0.4 (-)&  0.8 (-)&  1.0 (-)\\
		\emph{navigational} & 0.05 (0.05)& 0.3 (0.95)&  0.5 (-)&  0.7 (-)&  0.95 (-)\\
		\emph{informational} &  0.4 (0.3)&  0.6 (0.7)&  0.7 (-)&  0.8 (-)&  0.9 (-)\\
		\hline\hline
		& \multicolumn{5}{c}{$P(\mathit{stop}=1\mid click=1,  rel(d))$} \\
		%\hline
		\cmidrule(r){2-6}
		\emph{rel(d)} & 0 & 1 & 2 & 3 & 4 \\
		\hline
		\emph{perfect}  & 0.0 (0.0)& 0.0 (0.0)&  0.0 (-)&  0.0 (-)&  0.0 (-)\\
		\emph{navigational} & 0.2 (0.2)&  0.3  (0.9)&  0.5 (-)&  0.7 (-)&  0.9 (-)\\
		\emph{informational} & 0.1 (0.1)&  0.2 (0.5)&  0.3 (-)&  0.4 (-)&  0.5 (-)\\
		\hline
	\end{tabular}
	\label{tbl:sdbn}
\end{table}

\end{document}

%% file: sections/introduction.tex
\section{Introduction}
Online learning to rank  (OLTR)~\cite{hofmann2013fast,oosterhuis2018differentiable,zhuang2020counterfactual,oosterhuis2021unifying} aims to learn effective rankers from users search interactions, i.e., queries and clicks on search engine result pages (SERPs), by iteratively training and updating a production ranker through online interventions. The use of clicks, rather than relevance labels, reduces the high cost and time required to collect labels from editorial teams; it also better aligns with the user's true preferences than labels provided by third-party judges. The execution of this training process online rather than offline (e.g., as in counterfactual LTR~\cite{jagerman2019model}) addresses issues associated with rapid changes in query intents~\cite{zhuang2021how}.

Traditional OLTR solutions assume the ranker resides on a central server that controls the production of SERPs, including the online intervention made to explore the ranker's parameter space based on the index and that logs every user interaction (queries, clicks). This architecture, however, is inadequate for search contexts where the data is private or confidential and cannot be shared with the central search service or where users demand their interactions to be private, i.e. not to share clicks on SERPs with the server. Federated OLTR~\cite{kharitonov2019federated,wang2021effective} (FOLTR) has been canvassed as a solution to such situations. In FOLTR, private user data is kept on the user's device. The data is used locally within the user device to learn updates to a globally shared ranker. Local updates from all clients in the federated system are then shared to a central server\footnote{We note that while the use of a single central server is common among federated learning methods (and certainly is the only setup investigated so far for FOLTR), alternative setups are possible and include peer-to-peer federated systems with no central servers~\cite{lalitha2019peer,roy2019braintorrent,wang2021non}, and federated systems with multiple central servers.} (thus without sharing of actual user data),  which is responsible for the aggregation of the local updates, the consequent update of the global model and the sharing of the new global model with the clients (see Figure~\ref{fig:FOLTR} for a concrete example of a FOLTR system). The object of FOLTR is to federatively create a ranker that is more effective than each of the individual rankers users could create on each of the users private data -- and ideally this federated ranker should perform as well as a ranker that is created using all user data in a centralised manner.

Research on the effectiveness of the FOLTR paradigm and the factors that affect its performance is still limited to date, with only a couple of proposed and empirically investigated methods~\cite{kharitonov2019federated,wang2021federated,wang2021effective}. Importantly, research on FOLTR has fully ignored a key issue affecting the performance of federated learning (FL) systems: the presence of bias in how the training data is divided across the clients that join the federation. In other words, the fact that clients may hold non-independent and identically distributed (non-IID) data~\cite{zhu2021federated}.

\begin{figure}[t]
	\centering
	%\vspace{-10pt}
	\includegraphics[width=0.95\columnwidth]{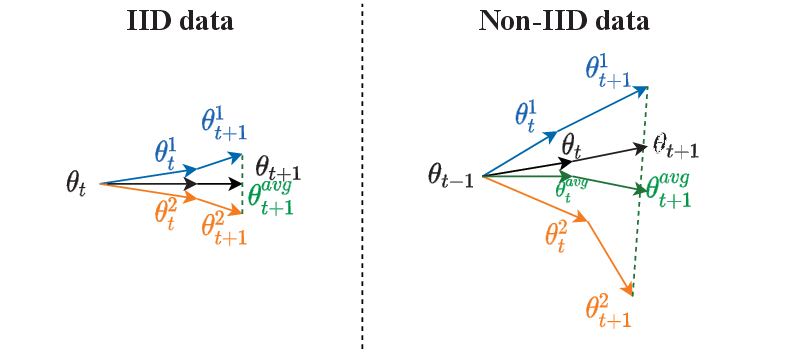}
	%\vspace{-6pt}
	\caption{Illustration of the model divergence problem in FL, adapted from~\citet{zhu2021federated}. $\theta_t$ is the ideal global model under centralised learning, and $\theta^{avg}_t$ is the average model created from the local models of Client 1 ($\theta^1_t$) and Client 2 ($\theta^2_t$) through FedAvg~\cite{mcmahan2017communication}.}
	\label{fig:modeldiver} 
\end{figure}

Non-IID data can pose severe threat to the effectiveness of a federated learning method. Models trained federatively in the presence of non-IID data across the clients that participate in the federation, in fact, display significantly lower effectiveness, and at times experience difficulties for the model to converge~\cite{zhu2021federated}. 
Effectiveness degradation can mainly be attributed to the weight divergence between the local models resulting from the non-IID distribution of the data across the clients~\cite{zhu2021federated}. Local models with the same initial parameters will converge to different models because of the heterogeneity of the local data distributions. This divergence will increase as more communication rounds of the federated learning algorithm are performed. This slows down or even impedes model convergence, worsening the performance of the global model. An illustration of the phenomenon of model divergence for both IID and non-IID data in federated learning is given in Figure~\ref{fig:modeldiver}. The ideal global model ($\theta_t$, under centralised learning) and actual global model ($\theta^{avg}_t$, average model created through FedAvg~\cite{mcmahan2017communication}) coincide when data is IID, but diverge when data is non-IID, showing that this is a sizeable problem when the data is non-IID.

This perspective paper $\footnote{In this paper, if not specified otherwise, we only consider horizontal FL~\cite{yang2019federated} and we believe our framework can be applied to both cross-device and cross-silo federated learning~\cite{kairouz2021advances}.}$ provides a systematic understanding of when non-IID data may occur in the FOLTR setting and the impact of non-IID data in such cases. This crucial research sheds light on the factors that need to be considered when devising and deploying FOLTR methods. It also details the experimental conditions for simulating non-IID data in FOLTR, paving the way for the development and adaptation to OLTR of existing and new methods for dealing with non-IID data. With this regard, we also show how some of the methods proposed in the federated learning literature to deal with non-IID data can be cast in the FOLTR framework and the gaps that still exist in effectively addressing non-IID data in FOLTR.

%% file: sections/related_works.tex
\section{Related Work}

\subsection{Federated Learning with non-IID data}
Zhu et al. have compiled a comprehensive survey on the impact of non-IID data on federated learning ~\cite{zhu2021federated}, also reviewing the current research on handling these challenges.  
Early work from Zhao et al.~\cite{zhao2018federated} shows a deterioration of the accuracy of federated learning if non-IID or heterogeneous data is present; they also provide a solution to this problem by creating a small subset of globally shared data between all clients (local devices).
Li et al.~\cite{li2020convergence} analyse the convergence of the federated learning algorithm \verb|FedAvg|~\cite{mcmahan2017communication} (which is a component of the FOLTR method we rely upon for investigation~\cite{wang2021effective}) on non-IID data and empirically show that data heterogeneity slows down the convergence. This raised attention to the presence of non-IID data in federated learning. 

Generally speaking, existing approaches for handling non-IID issues in federated learning can be classified into three categories: data-based approaches, algorithm-based approaches, and system-based approaches~\cite{zhu2021federated}.
Data sharing~\cite{zhao2018federated} and data augmentation~\cite{duan2019astraea} are two kinds of typical data-based approaches. While they achieve state-of-the-art performance, they fundamentally conflict with the objective of federated learning: that of not sharing data across clients. This is because, for example, methods such as data sharing require a subset of private data to be shared across all clients. While proposals have been made to use synthetic, rather than real, data for the data sharing mechanism~\cite{wang2021non} it is unclear (1) what the effectiveness loss of the sharing of synthetic data in place of real data is, and (2) whether the sharing of synthetic data could still jeopardise privacy as this synthetic data is typically generated from real data, and thus analysis of the synthetic data may reveal key aspects of and information contained in the real data.
Algorithm-based approaches mainly focus on personalisation methods like local fine-tuning of a neural model~\cite{wang2019federated} and Personalized FedAvg (Per-FedAvg)~\cite{fallah2020personalized} -- which are both limited mainly to neural models -- or the casting of the federated learning process into a multi-task learning problem~\cite{smith2017federated}.
System based approaches adopt clustering~\cite{sattler2020clustered} and tree-based structure~\cite{ghosh2020efficient} to deal with non-IID data.
Limitations exist among all proposed approaches, and this is still a much unexplored line of research.

\subsection{Federated Learning in IR}
We provide an overview of the use of federated learning in OLTR in section 3. That section also introduces the FOLTR method used in the empirical experimentation in this paper: the Federated Pairwise Differentiable Gradient Descent (FPDGD) method~\cite{wang2021effective}, which is the current state-of-the-art in FOLTR.

Aside from its usage in OLTR,  recent works have applied federated learning in other IR contexts. Zong et al.~\cite{zong2021fedcmr} provide a solution for cross-modal retrieval in a distributed data storage scenario, which uses federated learning to reduce the potential privacy risks and the high maintenance costs encountered when dealing with a large amount of training data. Wang et al.~\cite{wang2021efficient} study learning to rank (but not OLTR) in a cross-silo federated learning setting; this work is aimed at helping companies that have access to limited labelled data to collaboratively build a document retrieval system efficiently.
Hartmann et al.~\cite{hartmann2019federated} use federated learning to improve the ranking of suggestions
in the Firefox URL bar, so that the training of the ranker on user interactions is performed in a privacy-preserving way; they show that this federated approach improves on the suggestions produced by the previously employed heuristics in Firefox.
Yang et al. ~\cite{yang2018applied} describe the use of federated learning for search query suggestions in the Google Virtual Keyboard (GBoard) product. Here, a baseline model identifies relevant query suggestions given a user query; candidate suggestions are then filtered using a triggering model learnt using federated learning. 
Closest to FOLTR is the work of ~\citet{li2021federated}, who devise an offline federated learning method for counterfactual learning to rank from historic click logs. 

Aside from the previous examples, federated learning has also seen adoption in the area of personalised search~\cite{ghorab2013personalised}, which aims to return search results that cater to the specific user's interests. While feature-based~\cite{carman2010towards, bennett2012modeling, harvey2013building} and deep learning-based~\cite{song2014adapting, ge2018personalizing, yao2020rlper} methods are widely used in this area, user data privacy has been often overlooked -- this is particularly the case when considering the user's query logs which are collected by the central server to create the personalised ranker. To tackle this issue, Yao et al.~\cite{yao2021fedps} recently proposed a privacy protection enhanced personalised search framework which adapts federated learning to the state-of-the-art personalised search model. While not directly related to the OLTR context we consider here, these related lines of research could benefit from the investigations and considerations reported in this paper, as the problem of non-IID data in these previous contexts has also been ignored.

%% file: sections/foltr.tex
\begin{figure}[t]
	\centering
	%\vspace{-10pt}
	\includegraphics[width=0.95\columnwidth]{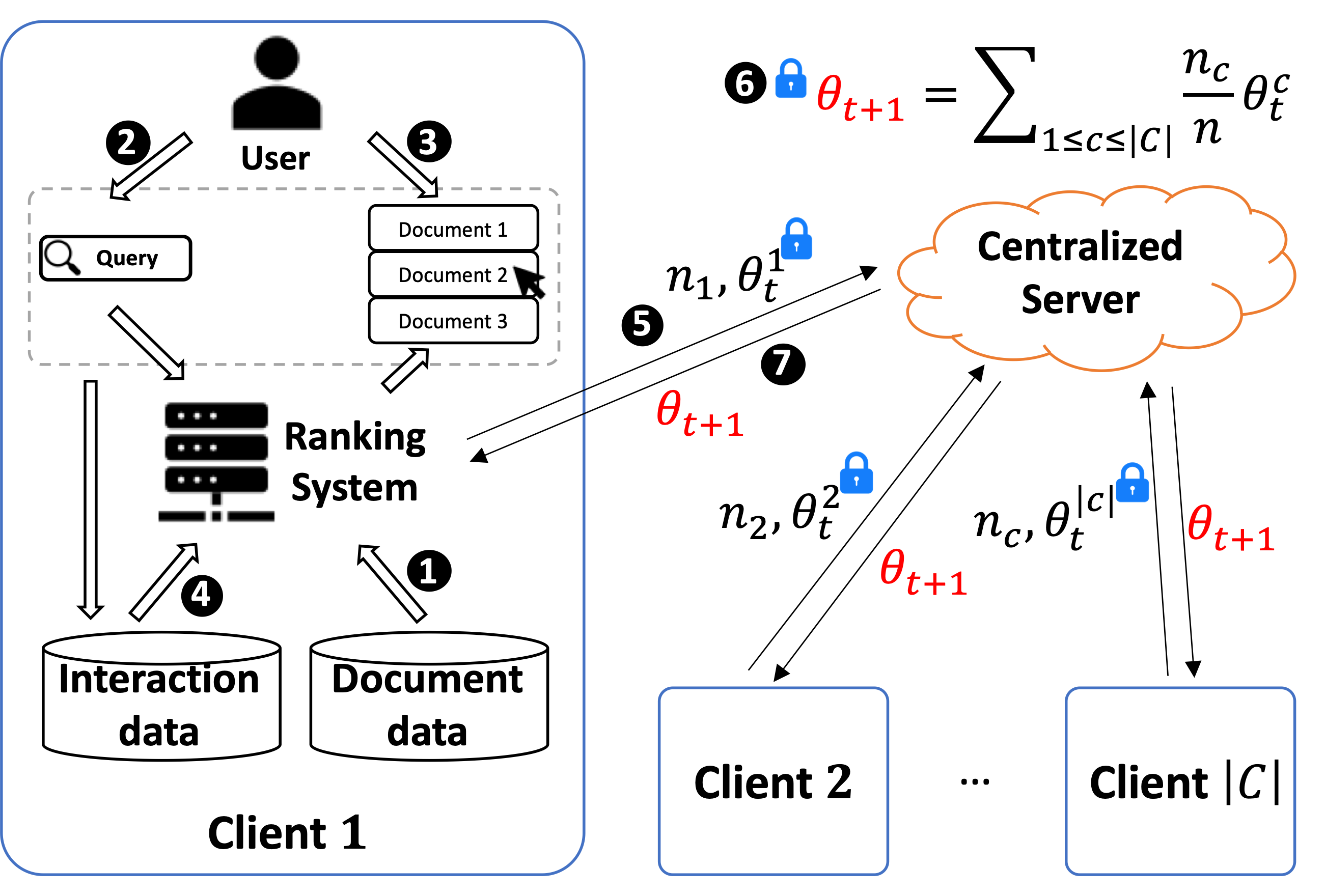}
	%\vspace{-6pt}
	\caption{Schematic representation of the FOLTR setting.
		\label{fig:FOLTR} %\vspace{-10pt}
	}
\end{figure}

\section{FOLTR Framework and FPDGD}

We next briefly describe the FOLTR framework, including the Federated Pairwise Differentiable Gradient Descent (FPDGD) method~\cite{wang2021effective}, which represents the current state-of-the-art in FOLTR and that we use as a representative method in our experiments to investigate the effect of non-IID data on FOLTR.

The federated online learning to rank setting is pictured in Figure~\ref{fig:FOLTR}. Searchable data is stored by each client (\circled{1}) and not shared with the centralised server or other clients. Different clients may hold all, a portion of, none of the same searchable data. Queries and user's clicks occur at a client side (\circled{2} and \circled{3}) and are not communicated to the centralised server or other clients: search is indeed entirely performed on the user device (\circled{2}). Each client exploits search interactions to perform local model updates to the ranker; for FPDGD, the routine executed by the client is shown in Algorithm~\ref{alg:fpdgd}, and the PDGD update is shown in Algorithm~\ref{alg:pdgd}. Each client considers $B$ interactions before updating the local ranker using the PDGD gradients. 
These local updates are then shared with the central server (\circled{5}), which in turn combines the ranker updates from the clients to produce a revised ranker (\circled{6}); for FPDGD, this is achieved according to the server routine in Algorithm~\ref{alg:fpdgd}. The new global model is then distributed to the user's device (\circled{7}).

\begin{algorithm}[t]
	\caption{FederatedAveraging PDGD. \\
		$\bullet$ set of clients participating training: $C$, each client is indexed by $c$; \\
		$\bullet$ number of local interactions for client $c$: $n_c$ ($\sum_{c=1}^{|C|}n_c=n$) \\
		$\bullet$ local interaction set: $B$, model weights: $\theta$.}
	\label{alg:fpdgd}
	\begin{algorithmic}
		\SUB{Server executes:}
		\STATE initialize $\theta_0$; scoring function: $f$; learning rate: $\eta$
		\FOR{each round $t = 1, \dots \infty$}
		\FOR{each client $c \in C$ \textbf{in parallel}}
		\STATE $\theta_{t+1}^c, n_c \leftarrow \text{ClientUpdate}(c, \theta_t)$
		\ENDFOR
		\STATE $\theta_{t+1} \leftarrow \sum_{c=1}^{|C|} \frac{n_c}{n} \theta_{t+1}^c$
		\ENDFOR
		\STATE
		
		\SUB{ClientUpdate($c, \theta_{t}$):}\ \ \ // \emph{Run on client $c$}
		\FOR{each local update $i$ from $1$ to $B$}
		\STATE $\theta_{t+1}^c \leftarrow \theta_{t} + \eta\nabla f_{\theta_{t}}^c$    \hfill \textit{\small //PDGD update shown in Algorithm.~\ref{alg:pdgd}}
		\ENDFOR
		\STATE return ($\theta_{t+1}^c, n_c$) to server
	\end{algorithmic}	
\end{algorithm}

\begin{algorithm}[t]
	\caption{Pairwise Differentiable Gradient Descent(PDGD)~\cite{oosterhuis2018differentiable}} 
	\label{alg:pdgd}
	\begin{algorithmic}[1]
		\STATE \textbf{Input}: initial weights: $\mathbf{\theta}_1$; scoring function: $f$; learning rate $\eta$.  \label{line:novel:initmodel}
		\FOR{$t \leftarrow  1, \ldots B$ }
		\STATE $q_t \leftarrow \mathit{receive\_query}(t)$\hfill \textit{\small // obtain a query from a user} \label{line:novel:query}
		\STATE $D_t \leftarrow \mathit{preselect\_documents}(q_t)$\hfill \textit{\small // preselect documents for query} \label{line:novel:preselect}
		\STATE $\mathbf{R}_t \leftarrow \mathit{sample\_list}(f_{\theta_t}, D_t)$ \hfill \textit{\small // sample list} \label{line:novel:samplelist}
		\STATE $\mathbf{c}_t \leftarrow \mathit{receive\_clicks}(\mathbf{R}_t)$ \hfill \textit{\small // show result list to the user} \label{line:novel:clicks}
		\STATE $\nabla f_{\theta_{t}} \leftarrow \mathbf{0}$ \hfill \textit{\small // initialize gradient} \label{line:novel:initgrad}
		\FOR{$d_k >_{\mathbf{c}} d_l \in \mathbf{c}_t$} \label{line:novel:prefinfer}
		\STATE $w \leftarrow \rho(d_k, d_l, R, D)$  \hfill \textit{\small // initialize pair weight} \label{line:novel:initpair}
		\STATE $w \leftarrow w 
		\frac{ 
			e^{f_{\theta_t}(\mathbf{d}_k)}e^{f_{\theta_t}(\mathbf{d}_l)}
		}{
			(e^{f_{\theta_t}(\mathbf{d}_k)} + e^{f_{\theta_t}(\mathbf{d}_l)}) ^ 2
		}$
		\hfill \textit{\small // pair gradient} \label{line:novel:pairgrad}
		\STATE  $\nabla f_{\theta_{t}} \leftarrow \nabla f_{\theta_{t}} + w (f'_{\theta_t}(\mathbf{d}_k) - f'_{\theta_t}(\mathbf{d}_l))$
		\hfill \textit{\small // model gradient} \label{line:novel:modelgrad}
		\ENDFOR
		\STATE $\theta_{t+1} \leftarrow \theta_{t} + \eta \nabla f_{\theta_{t}}$
		\hfill \textit{\small // update the ranking model} \label{line:novel:update}
		\ENDFOR
	\end{algorithmic}
\end{algorithm}

%% file: sections/noniid_data.tex
\section{Types of non-IID Data in FOLTR}

We consider training a ranker for the OLTR system as a supervised learning task in an FL setup, with each client holding a subset of the data. Each data sample is denoted as $(x,y)$, where $x$ is the feature representation of the data and $y$ is the label. The local distribution of the dataset in client $i$ is denoted as $P_i(x,y)$. The presence of non-IID data can be represented as the difference between local data distributions: that is, for different clients $i$ and $j$, $P_i(x,y) \neq P_j(x,y)$.

In federated learning, data across clients may not be IID due to different reasons: Kairouz et al.~\cite{kairouz2021advances} and Zhu et al.~\cite{zhu2021federated} assert this can be due to how features $x$ and labels $y$ are distributed. However, the translation of these categories to FOLTR is not straightforward. In the following sections, we put forward several situations in which data specific to FOLTR could be distributed in a non-IID manner across clients. Specifically, we consider data in the FOLTR process may not be IID because of biases across clients due to:
\begin{itemize}
	\item \textbf{Type 1:} document preferences (Section~\ref{sec:type1})
	\item \textbf{Type 2:} document label distribution skewness (Section~\ref{sec:type2})
	\item \textbf{Type 3:} click preferences (Section~\ref{sec-type3-4})
	\item \textbf{Type 4:} data quantity (Section~\ref{sec-type3-4})
\end{itemize}

\begin{table*}[t]
	\newcommand{\tabincell}[2]{\begin{tabular}{@{}#1@{}}#2\end{tabular}}

	\centering
	\caption{Summary of non-IID data types in FOLTR. \label{tbl-data-types}}
	\begin{tabular}{cll}
		\hline
		\bf Data type & \bf Key characteristic   & \bf When it happens in FOLTR                                                                                        \\ \hline
		Type 1    & Document Preferences & \tabincell{l}{Different clients have different preferred candidate documents, although they are \\searching for the same query.} \\ \hline
		Type 2    & Document Label Distribution & \tabincell{l}{Different clients hold candidate documents with different label distribution while \\the conditional feature distirbution is the shared.} \\ \hline
		Type 3    & Click Preferences & \tabincell{l}{Different clients have various preferred click behaviours when searching for the \\same query.} \\ \hline
		Type 4    & Data Quantity & \tabincell{l}{Different clients have different frequency on issuing queries and interacting with \\the searching system.} \\ \hline
	\end{tabular}
	
\end{table*}

The last data type, Type 4, i.e., the situation in which different clients hold different quantities of data (and in particular interaction data such as queries and clicks), does not necessarily imply that the data is non-IID. However, we note this case is often studied in the FL literature alongside non-IID data~\cite{zhu2021federated,li2022federated}, and thus we include this situation in our considerations of the non-IID problem. Each data type is defined and investigated in the next sections; in addition we provide a summary overview of the data types in Table~\ref{tbl-data-types}.

We also note that commonly in federated learning, non-IID data occurs because the data is distributed across clients according to its features. In other words, the marginal distribution of the features belonging to the data held by each client may vary, i.e. for different clients $i$ and $j$, $P_i(x) \neq P_j(x)$. This situation may occur in horizontal federated learning settings (also called homogeneous FL)~\cite{yang2019federated}, where each client holds different and overlapping datasets. In this case, the non-IID divergence is usually caused by inconsistent data distributions, e.g., feature imbalance of the training data local to each client. However, this case does not seem applicable to FOLTR (thus is not further studied in this paper). In FOLTR, each data item is represented by the feature vector of a query-document pair and its relevance label. 
The features often consist of variations of query-dependent features such as TF-IDF scores, BM25 scores, query length, as well as query-independent features such as PageRank, URL lengths, and so on~\cite{qin2013introducing}. In this case, bias in the feature distribution across clients would be rare as most features are dependent on the query-document pair. 

Next, we describe the non-IID data types we put forward in this paper and analyse their impact on FOLTR. We empirically find that only Type 1 and partially also Type 2 data have a strong impact on the FOLTR. We thus predominantly focus our attention on these two data types while providing only a definition and a brief account of the remaining two data types in the paper due to space: we do, however, report all experiments results, thorough analysis and considerations in an online appendix available at \url{https://github.com/ielab/non-iid-foltr}.

\input{sections/type1.tex}

\input{sections/type2.tex}

\input{sections/type3-4.tex}

%% file: sections/type1.tex
\section{Type 1: Document preferences} \label{sec:type1}

\emph{Document preference skewness} (Type 1) considers the situation when the conditional distribution $P_i(y|x)$ varies across the clients though $P_i(x)$ remains the same. This happens when different clients have different preferred candidate documents, although they are searching for the same query. As OLTR requires the user's implicit feedback as an optimization objective, which might be highly related to individual preferences, this setting appears to be of very likely occurrence.

\begin{figure*}[t]
	\centering
	\includegraphics[width=17cm]{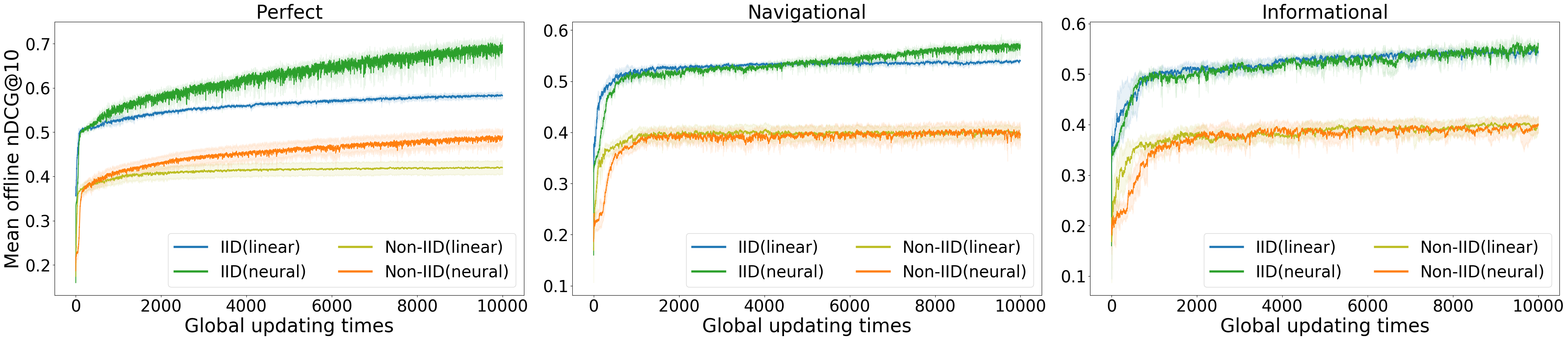}
	\vspace{-8pt}
	\caption{Offline performance (nDCG@10) on Type 1 data; results averaged across dataset splits and experimental runs.}
	\label{fig:dps}
\end{figure*}

\subsection{Simulating Type 1 non-IID Data}

The mechanism we use to simulate non-IID data of Type 1 and IID data to baseline the FOLTR effectiveness relies on a recent work that empirically studied and demonstrated how OLTR methods adapt when user's search intents change overtime~\cite{zhuang2021how}. In particular, \citet{zhuang2021how} created a collection for OLTR with several explicit intent types by adapting an existing TREC collection, as no dataset is available for studying this OLTR problem. Derived from ClueWeb09 and the TREC Web Track 2009 to 2012~\cite{clarke2012overview}, this intent change collection consists of 200 queries with 4 intents each and, on average, 512 candidate documents per query. Furthermore, query-document pairs' relevance judgements are provided per intent. We believe this is an appropriate collection to adapt to study the effect of Type 1 non-IID data on FOLTR. We can regard each intent as a type of user preference. As the average number of relevant documents per intent varies largely across all intent types, the learning difficulty of optimizing a ranker among different intents also varies. To avoid this bias, we follow Zhuang and Zuccon~\cite{zhuang2021how} and we re-label the original intent number for each query through random shuffling: this is possible because all intent types are independent across queries. In our experiments, we repeat this process of re-balancing 5 times, thus giving rise to results averaged across 5 FOLTR experiments. We refer to Zhuang and Zuccon~\cite{zhuang2021how} for further details on the dataset creation, and we further highlight that we have made available an implementation of the dataset creation procedure along with the actual dataset at \url{https://github.com/ielab/non-iid-foltr}.

To simulate non-IID data, after randomly shuffling all intents across 4 types, we let each intent represent one client preference. The client preferences differ from each other for the same query-document pair so as the corresponding relevance judgements. The federated setup involves 4 clients (represented by 4 types of intent) and the local updating time $B = 5$ with fixed global communication times $T = 10,000$. These settings are similar to those used in previous work on FOLTR~\cite{kharitonov2019federated,wang2021federated,wang2021effective} -- in particular we refer the interested reader to the work of Wang et al.~\cite{wang2021federated} to understand the relationships between number of clients, number of local updates $B$, and FOLTR effectiveness.
For the implicit feedback in FOLTR, we simulate user clicks based on the popular \emph{Simplified Dynamic Bayesian Network} (SDBN) click mode~\cite{chapelle2009dynamic}, following settings in previous work on OLTR~\cite{oosterhuis2016probabilistic,oosterhuis2018differentiable,zhuang2020counterfactual,wang2021effective}. We limit SERP to 10 documents and use $nDCG@10$ for offline evaluation, cumulative discounted $nDCG@10$~\cite{oosterhuis2018differentiable} for online evaluation. We train a linear ranker and a neural ranker on the intent-change dataset. As in \citet{zhuang2021how}, given that no held-out test set is available, we evaluate both online and offline performance on the original training set across all 4 intent types and average all results. For the IID setting, we merge all intents and mark a document as relevant as long as it is judged relevant for at least one of the intent types. Each client randomly picks a query from the training set and clicks documents based on the same preferences during the federated training with IID data. Other settings remain the same as the non-IID experiments.

\subsection{Impact of Type 1 non-IID Data}

The offline performance related to Type 1 data is shown in Figure~\ref{fig:dps}; the corresponding online performance is shown in Table~\ref{tbl:dps_online}.
From the offline performance, it is clear that the presence of non-IID data negatively impacts the performance of the learnt ranker, compared to those obtained when data is IID. In terms of online performance, rankers obtained in the presence of non-IID data are also worse than when trained with IID data. This can be explained as follows. Since each client has its preference (intent), the relevant documents are judged in different ways; this leads to the divergence of each client's local ranker update, as exemplified in Figure~\ref{fig:modeldiver}.

In summary, we find that if data is distributed in a non-IID manner across clients according to Type 1, the effectiveness of FOLTR (and specifically of FPDGD) is seriously affected.

\begin{table}[t]
	\centering
	\caption{Online performance on Type 1 data, averaged across dataset splits and experimental runs. Significant differences between IID and non-IID are indicated by \dubbelop \ (p $<$ 0.05)}
	\begin{tabular}{ll ccc}
		\hline
		ranker&data types&\emph{perfect}&\emph{navigational}&\emph{informational} \\
		\hline
		\emph{linear}&IID& 1002.36 {\tiny \dubbelop} & 872.12 {\tiny \dubbelop} & 894.95 {\tiny \dubbelop} \\
		&non-IID& 648.71 & 546.25 & 566.23 \\
		\hline
		\emph{neural}&IID& 1061.57 {\tiny \dubbelop} & 834.08 {\tiny \dubbelop} & 842.87 {\tiny \dubbelop} \\
		&non-IID& 668.38 & 505.64 & 490.29 \\
		\hline
	\end{tabular}
	\label{tbl:dps_online}
\end{table}

\subsection{Dealing with Type 1 non-IID Data} \label{sec:type1dealing}
The employed state-of-the-art FPDGD method is based on the FedAvg algorithm. The fact that FPDGD is affected by non-IID data may be due to the underlying federation algorithm, i.e. FedAvg itself. In federated learning literature, variations of this federation algorithm have been proposed to tackle the non-IID data problem directly. We select two of such methods, FedProx~\cite{li2020federated} and FedPer~\cite{arivazhagan2019federated}, and adapt them to the FPDGD method.

\begin{figure*}[t]
	\centering
	\subfigure[\textbf{intent-change (linear ranker) - FedProx}] {\label{fig:intent-linear-fedprox} \includegraphics[width=17cm]{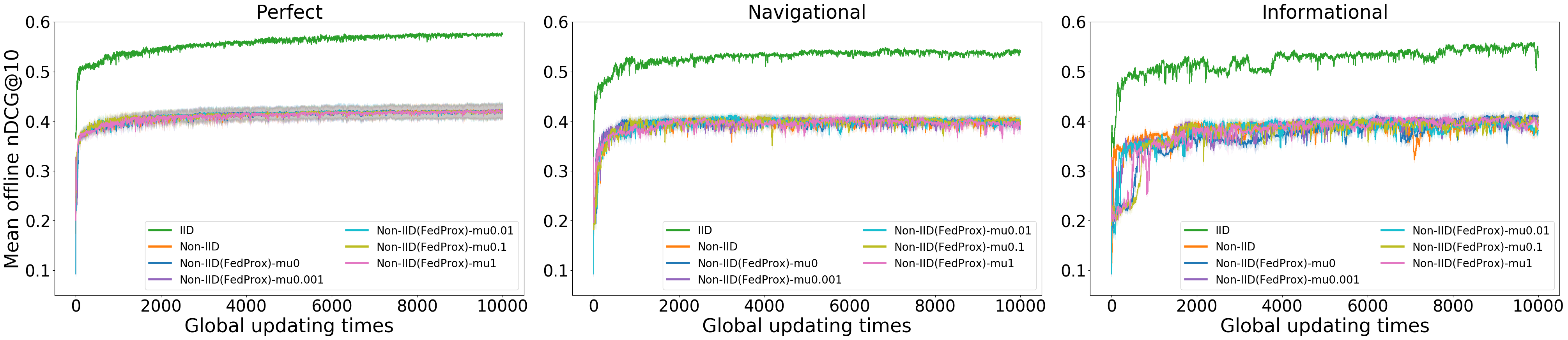}}
	\subfigure[\textbf{intent-change (neural ranker) - FedProx}] {\label{fig:intent-neural-fedprox} \includegraphics[width=17cm]{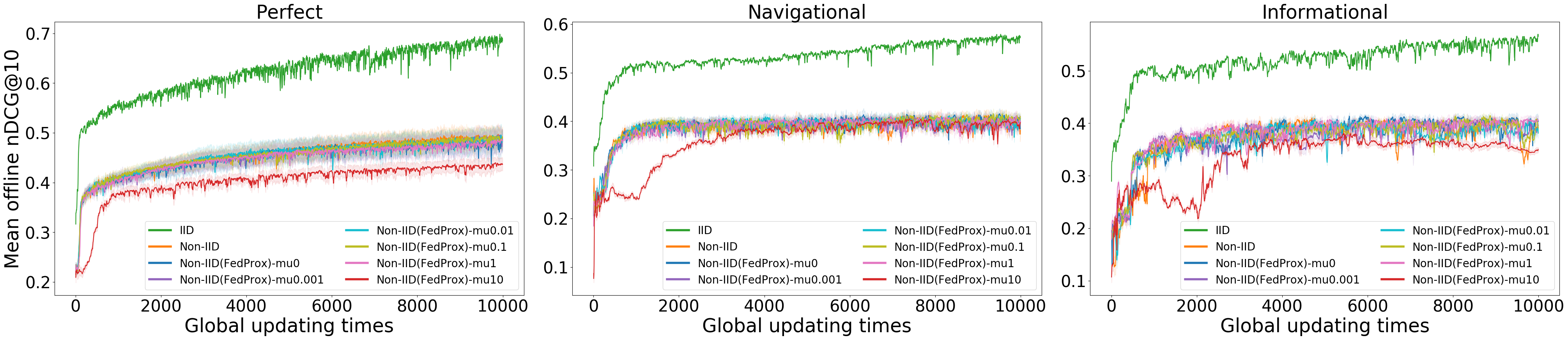}}
	\subfigure[\textbf{intent-change (neural ranker) - FedPer}] {\label{fig:intent-neural-fedper} \includegraphics[width=17cm]{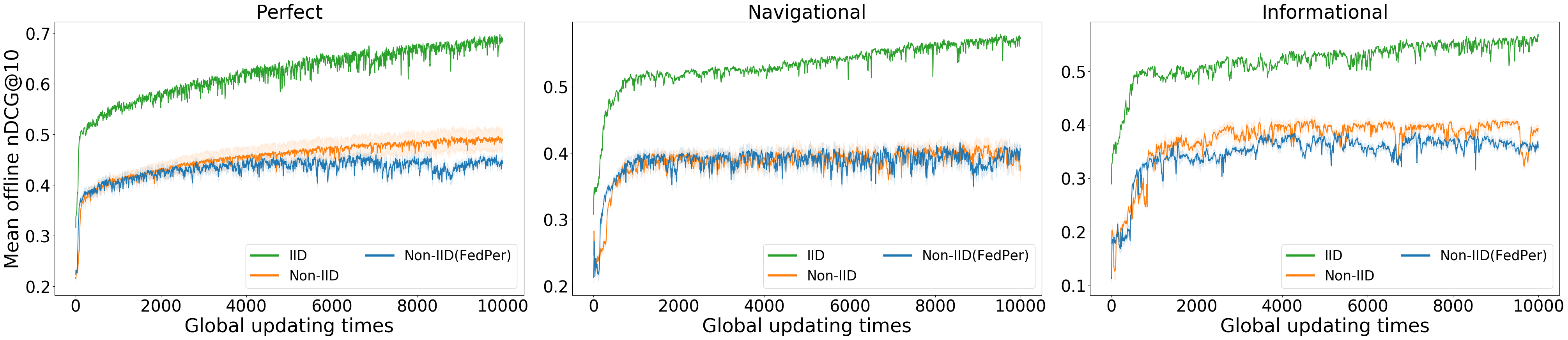}}
	\vspace{-12pt}
	\caption{Offline performance on Type 1 data for FedProx and FedPer; results averaged across dataset splits and experimental runs.
		\label{fig:dps-methods}}
\end{figure*}

FedProx~\cite{li2020federated} improves the local objective of FedAvg. Specifically, it introduces an additional $L_2$ regularisation term (weighted according to a hyper-parameter $\mu$) in the local objective function to limit the distance between the local model and the global model. We provide details of our adaptation of FedProx to FPDGD in the online appendix; the use of FedProx adds little computational overhead. However, the main drawback is that the hyper-parameter $\mu$ needs to be carefully tuned: a large $\mu$ may slow the convergence by forcing the updates to get close to the initial point, while a small $\mu$ may not make much difference compared to the use of FedAvg.

FedPer~\cite{arivazhagan2019federated} tackles the presence of non-IID exclusively for deep neural networks by separating them into base layers and personalisation layers.
The base layers are trained collaboratively through FedAvg, where all clients share the same base layers. Instead, the personalisation layers are trained locally using the clients' local data with stochastic gradient descent (SGD). This procedure works as follows: after initialisation, each client merges and updates its base and personalised layers locally using an SGD style algorithm. Each client only sends its base layers to the global server. The server updates the globally-shared base layers using FedAvg and sends back again the updated ones to each client. Intuitively, the base layers are updated globally to learn common high-level representations. In contrast, the distinct personalisation layers never leave the local device and capture the personalisation aspects required by the clients. Except for the training and the maintenance of the local personalisation layers, FedPer is quite similar to FedAvg. FedPer, however, reduces the communication costs as only part of the whole model is transferred and has shown enhanced learning performance under highly skewed non-IID data~\cite{arivazhagan2019federated}.

Our experimental results on FedProx and FedPer are shown in Figure~\ref{fig:dps-methods}; for FedProx we explored $\mu \in \{0.001, 0.01, 0.1, 1, 10\}$. The results clearly show that these federated learning methods, which successfully deal with non-IID data in general machine learning tasks, are not effective in the FOLTR context. In fact, not only do these methods not overcome the gap in effectiveness between IID and non-IID setups, but they even only provide limited improvements, if any, compared to FPDGD with FedAvg. This is an important finding because: (1) it shows a realistic case in which non-IID data largely affects FOLTR effectiveness, and (2) it shows that current methods developed in general FL for non-IID data do not work in FOLTR. Thus, a strong need for new methods specialised in the FOLTR settings emerges from these findings.

%% file: sections/type2.tex
\section{TYPE 2: Document label distribution skewness} \label{sec:type2}

\emph{Document label distribution skewness} (Type 2) is a widely recognised type of non-IID data type in federated learning. In this setting, the label distributions $P_i(y)$ in each client are different while the conditional feature distribution $P_i(x|y)$ is shared across the clients. In terms of FOLTR, this is equivalent to the following situation. Assume a document is evaluated across the $r$-level relevance grades, from \emph{not relevant} (0) to \emph{perfectly relevant} ($r-1$); then the label distribution on each client is such that, for client $i$, the probability of holding documents with relevance label $k$ is $P_i(R=k) = p_{k}$, where $\sum_{k=0}^{r-1}p_k = 1$, $\forall k, p_k \in [0,1]$

In practice, this may be represented by a situation like the following. Several hospitals are collaboratively creating a FOLTR ranker for clinical-decision-support~\cite{roberts2015overview,roberts2016state}. Certain hospitals hold a significantly larger portion of highly relevant health records for a certain disease, while some only a small fraction. In this circumstance, the document label distribution is skewed. Under the context of email search~\cite{narang2017large}, different clients might have unique strategies for managing personal emails~\cite{whittaker1996email}. Some clients frequently clean up their inboxes and use folders to organise emails. In contrast, some hardly use folders or delete irrelevant messages, resulting in different label distribution when following a learning-to-rank approach.

\begin{figure*}[t]
	\centering
	\subfigure[\textbf{MSLR-WEB10k (linear ranker)}] { \label{fig:mslr-r1-linear} \includegraphics[width=17cm]{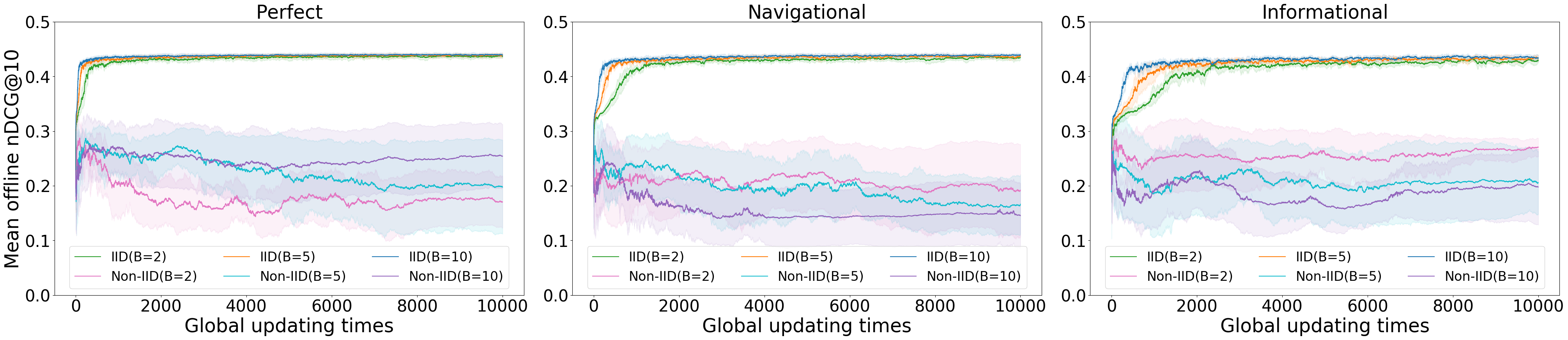}} 
	\subfigure[\textbf{MSLR-WEB10k (neural ranker)}] { \label{fig:mslr-r1-neural} \includegraphics[width=17cm]{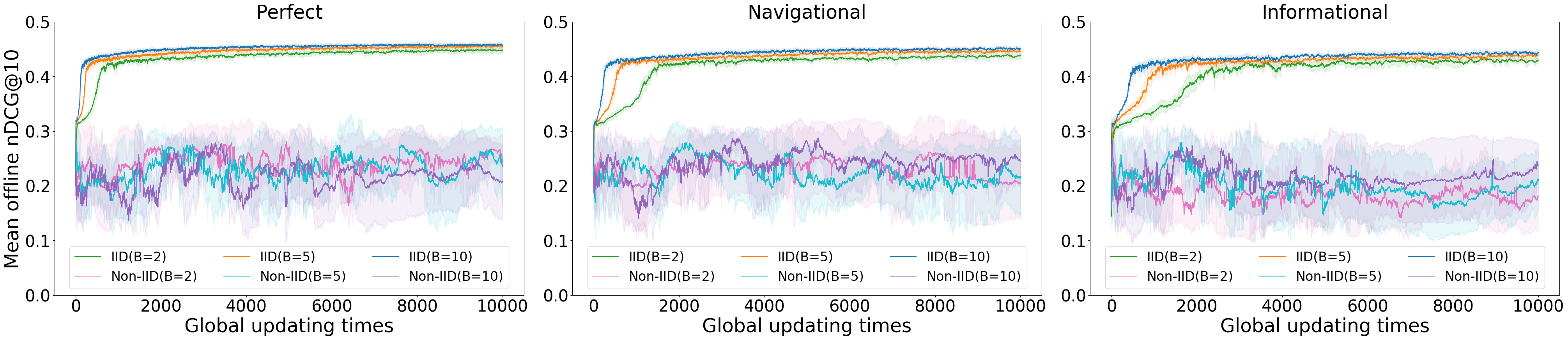}} 
	\vspace{-12pt}
	\caption{Offline performance (nDCG@10) on MSLR-WEB10k for Type 2 ($\#R = 1$), under three instantiations of SDBN click model and three local updates setting ($B \in \{2, 5, 10\} $); results averaged across all dataset splits and experimental runs.
		\label{fig:lds1}}
\end{figure*}

\begin{figure*}[t]
	\centering
	\subfigure[\textbf{MSLR-WEB10k (linear ranker)}] {\label{fig:mslr-linear} \includegraphics[width=17cm]{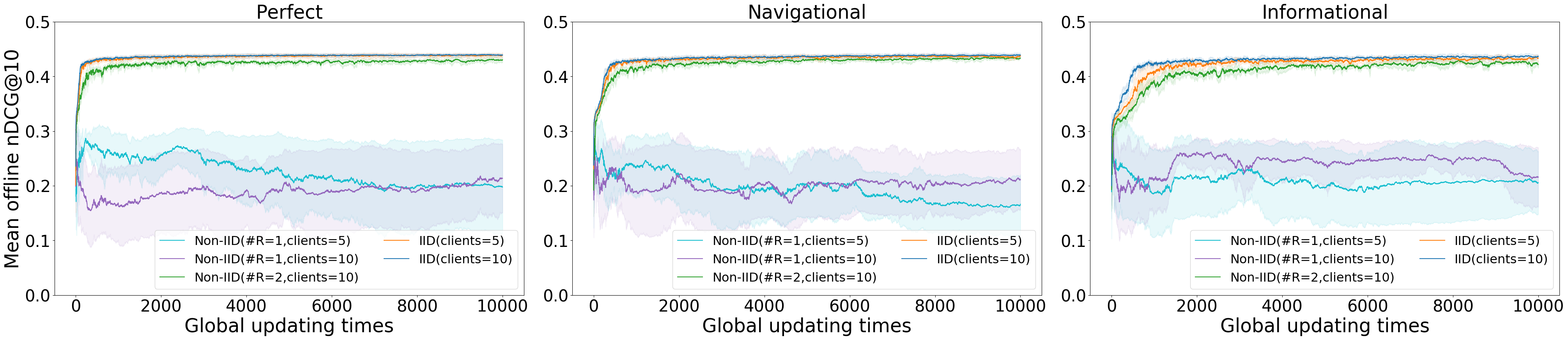}} 
	\subfigure[\textbf{MSLR-WEB10k (neural ranker)}] {\label{fig:mslr-neural} \includegraphics[width=17cm]{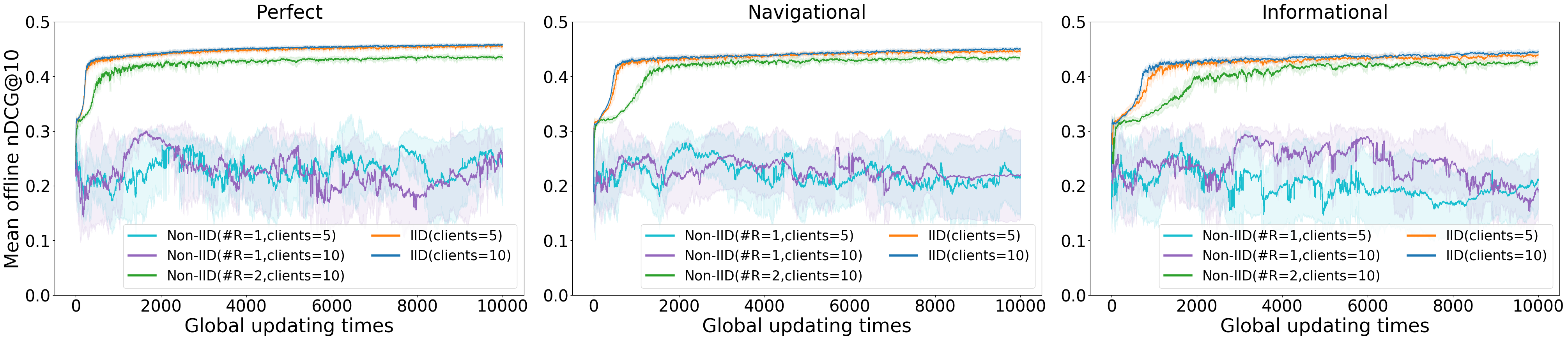}} 
	\vspace{-12pt}
	\caption{Offline performance (nDCG@10) on MSLR-WEB10k for Type 2 ($\#R = 2$), under three instantiations of SDBN click model with local updates setting ($B = 5$); results averaged across all dataset splits and experimental runs.
		\label{fig:lds2}}
\end{figure*}

\subsection{Simulating Type 2 non-IID Data}

In this section, we discuss how we synthetically simulate Type 2 non-IID data and IID data to baseline in the FOLTR effectiveness. For these experiments, we use the popular datasets MSLR-WEB10k~\cite{qin2013introducing} (10,000 queries), Yahoo~\cite{chapelle2011yahoo} (29,900 queries) and Istella-S~\cite{lucchese2016post} (33,018 queries). We report the results for MSLR-WEB10k in the paper; results on the other datasets are similar and are provided in the online appendix.
We simulate $|C|$ clients with each client performing $B$ interactions (queries) locally to contribute to each global model update and restrict the global communication times $T = 10,000$.
For simulating querying behaviour, for each client participating in the federated OLTR, we sample $B$ queries randomly, in line with previous work on FOLTR~\cite{kharitonov2019federated, wang2021effective}. For each query, we use the local ranking model (i.e. that held by the client) to rank documents; we limit SERP to 10 documents. 
For the click behaviour, we rely on the same SDBN click models as Section~\ref{sec:type1}. We train both a linear ranker and a neural ranker same as Wang et al.~\cite{wang2021effective}.

We specifically consider two types of non-IID data for Type 2: non-IID subtype 1 and non-IID subtype 2. The main difference between the two types is the number of different labels (i.e. the graded relevance assessments) in each client's local dataset. Following similar partitioning strategies by Li et al.~\cite{li2022federated}, suppose each client only has data samples for $k$ different labels. We first generate all possible $k$-combinations of the relevance set $R$ and randomly assign to $\binom{R}{k}$ clients. Then, for the query-document pairs of each label, we randomly and equally divide them into the clients who own the label. In this way, the number of labels in each client is fixed, and there is no overlap between the samples of different clients.

In non-IID subtype 1, each client only holds query-document pairs from one specific value of relevance label. We use $\#R = 1$ to denote this partitioning strategy. This federated setup involves $|C|=5$ clients. We also vary the local updating time $B \in \{2, 5, 10\}$ to investigate the impact of local updating with a fixed global communication time $T = 10,000$. For non-IID subtype 2, each client holds data samples from two relevance labels -- we denote this as $\#R = 2$. We simulate $|C|=10$ clients and for fair comparison between the two non-IID subtypes, we also simulate  $|C|=10$ clients for $\#R = 1$ (with each label distributed on two different clients).

The IID experimental setting is the same as the non-IID in terms of federation, ranker parameters and evaluation procedure, except that each client now randomly picks a query from the whole training set with all graded judgements during the training period.

\subsection{Impact of Type 2 non-IID Data}

The offline performance for $\#R = 1$ on MSLR-WEB10k is shown in Figure~\ref{fig:lds1} and the corresponding online performance is shown in Table~\ref{tbl:lds_online}. From the offline results, it is clear that the rankers learned with non-IID data under-fit the generalized held-out test set under all three settings of local updating times ($B$). For the \emph{perfect} click model, a larger number of $B$ achieves better test performance. However, when it comes to noisier clicks (\emph{navigational} or \emph{informational}), the trend is reversed, although differences are minimal and the model performance fluctuates. 
For the online results, all non-IID settings appear to over-fit the maximum value ($online\_ndcg = 1589.23$) as each client's local data only contains data from one relevance label.

\begin{table}[t]
	\footnotesize
	\centering
	\caption{Online performance on MSLR-WEB10k for Type 2 ($\#R = 1$), averaged across dataset splits and experimental runs.}
	\begin{tabular}{lccccccc}
		\hline
		& &\multicolumn{3}{c}{\emph{linear ranker}} & \multicolumn{3}{c}{\emph{neural ranker}}   \\
		\cmidrule(r){2-8}
		&click&$B = 2$&$B = 5$&$B = 10$&$B = 2$&$B = 5$&$B = 10$ \\
		\hline
		IID&\emph{per.}& 742.10 & 778.56 & 798.05 & 716.55 & 781.18 & 815.8 \\
		&\emph{nav.}& 698.25 & 743.35 & 771.04 & 649.83 & 728.96 & 775.87\\
		&\emph{inf.}& 672.23 & 722.23 & 757.10 & 612.76 & 693.35 & 748.64 \\
		\hline	
		non-IID&\emph{per.}& 1589.23 & 1589.23 & 1589.23 & 1589.23 & 1589.23 & 1589.23\\ 
		&\emph{nav.}& 1589.23 & 1589.23 & 1589.23 & 1589.23 & 1589.23 & 1589.23\\
		&\emph{inf.}& 1589.23 & 1589.23 & 1589.23 & 1589.23 & 1589.23 & 1589.23\\
		\hline
	\end{tabular}
	\label{tbl:lds_online}
	\vspace{-20pt}
\end{table}

\begin{figure*}[t]
	\centering
	\subfigure[\textbf{MSLR-WEB10k (linear ranker)}] {\label{fig:mslr-linear-datashare} \includegraphics[width=17cm]{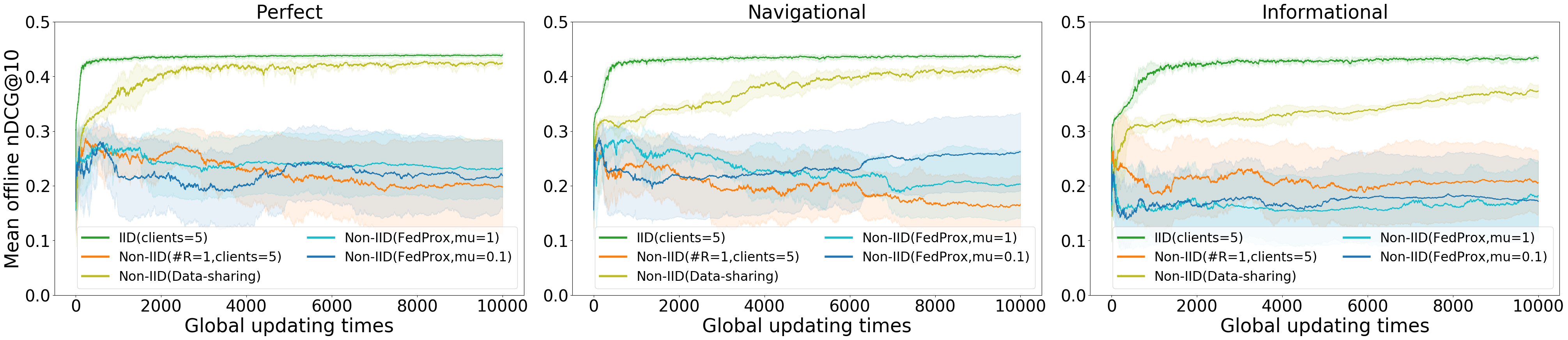}} 
	\subfigure[\textbf{MSLR-WEB10k (neural ranker) - data sharing}] {\label{fig:mslr-neural-datashare} \includegraphics[width=17cm]{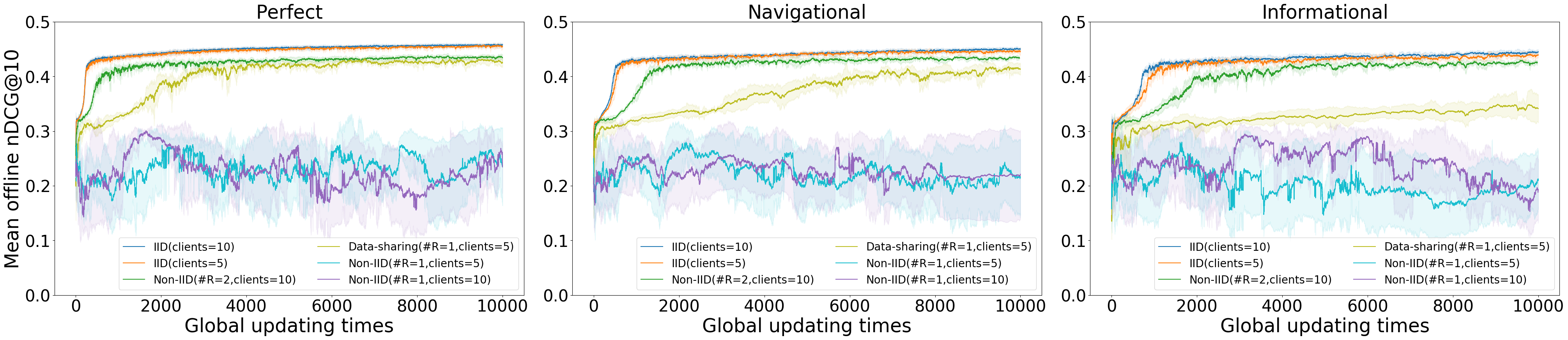}} 
	\subfigure[\textbf{MSLR-WEB10k (neural ranker) - FedProx \& FedPer}] {\label{fig:mslr-neural-fedper} \includegraphics[width=17cm]{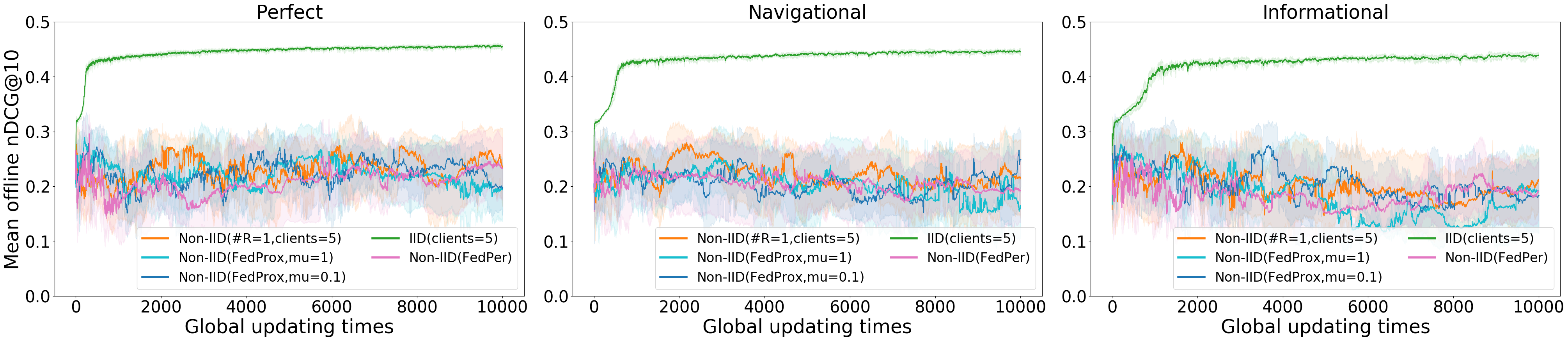}} 
	\vspace{-12pt}
	\caption{Offline performance on MSLR-10k when using Data-sharing, FedProx and FedPer on Type 2 non-IID data (with $\#R = 1$); results averaged across dataset splits and experimental runs.
		\label{fig:lds-methods}\vspace{-10pt}}
\end{figure*}

Offline results for $\#R = 2$ are shown in Figure~\ref{fig:lds2}. In this case, the effectiveness of the learnt rankers is much higher than for $\#R = 1$: a diversity in labels held by a client prevents major losses in FOLTR effectiveness. This result is also consistent with previous results in general federated learning with non-IID data: Li et al.~\cite{li2022federated} found that the most challenging setting is when each client only has data samples from a single class (label). We further note that another reason for the performance gap is the pairwise loss used in FPDGD~\cite{wang2021effective}: when each client only has one relevance label, it is hard to infer preferences between document pairs (as they both have the same label). However, given labels from two levels of relevance ($\#R = 2$), pairwise differences can be effectively inferred. This suggests that the results obtained here for Type 2 data may not generalise to other FOLTR methods beyond FPDGD if they do not rely on the pairwise preference mechanism. We further note, however, that FPDGD is the current state-of-the-art method and that the only available alternative~\cite{kharitonov2019federated} displays highly variable and sensibly worse performance compared to FPDGD~\cite{wang2021federated,wang2021efficient}. Therefore, new methods of FOLTR must also be validated in the presence of Type 2 data.

In summary, we find that if data is distributed in a non-IID manner across clients according to Type 2, the effectiveness of FOLTR (and specifically of FPDGD) is seriously affected in the case of $\#R = 1$; however, if $\#R = 2$ then gaps in effectiveness compared to IID settings are minimal. 

\vspace{-7pt}

\subsection{Dealing with Type 2 non-IID Data}

To mitigate the effect of Type 2 non-IID data, we investigate three existing methods from the federated learning literature: Data-sharing, FedProx and FedPer. FedProx and FedPer have been described in Section~\ref{sec:type1dealing}.
Data-sharing was first proposed by \citet{zhao2018federated}. They attribute the performance reduction observed on non-IID data to the weight divergence, which is further affected by the divergence between the local data distribution and the overall distribution. They then introduce a straightforward idea to improve FedAvg: slightly reduce the divergence that causes the global model to underperform. This can be achieved as follows. A globally shared dataset $G$ characterised by the overall data distribution is centralised on the server, and a warm-up global model is trained from $G$. Then, a random $\alpha$ proportion of $G$ is sent to all clients to update the local model by both local training data and the shared data from $G$. Lastly, the server aggregates the local models from the clients and updates the global model with FedAvg. Experimental results on machine learning tasks show that data sharing can significantly enhance the global model performance in the presence of non-IID data. However, the shortcomings are also pronounced. It is challenging to collect uniformly distributed global datasets in real-world scenarios because either the global server needs some prior knowledge about the local data distributions or each client needs to share parts of the local data (violating the privacy requirement underlying FL).

Figure~\ref{fig:lds-methods} reports the results for Data-sharing, FedProx and FedPer on MSLR-10k dataset under label distribution skewness $\#R = 1$. We randomly select 10\% of the entire dataset as the globally shared data and simulate $|C|=5$ clients with $B = 5$ local updates before each global update. 
Results show that the global performance can be significantly enhanced with data-sharing for both linear and neural rankers. On the other hand, neither FedProx nor FedPer provides statistically significant gains over the basic FPDGD on Type 2 non-IID data (with $\#R = 1$).

%% file: sections/type3-4.tex
\section{Other Data Types}
\label{sec-type3-4}

\subsection{Type 3: Click Preferences} 
Next, we consider as a source of non-IID data the noise and biases caused by the different click preferences arising from different clients that participate in the FOLTR training; we term this type of non-IID data as \emph{click preference skewness} (Type 3). 

The mechanism to emulate non-IID data of Type 3 and IID data baseline in our FOLTR experiments is as follows.
We study two widely used click models: the \emph{Simplified Dynamic Bayesian Network} (SDBN) click model~\cite{chapelle2009dynamic} and the \emph{Position-Based Model} (PBM)~\cite{craswell2008experimental}. For non-IID settings in SDBN, each client chooses one of three widely-used instantiations of SDBN, namely \emph{perfect, navigational, informational}. For the non-IID settings with PBM, we generate 5 instantiations based on varying the $\eta \in \{0, 0.5, 1, 1.5, 2\}$ parameter: each client is represented by one click type. Thus the federated setup involves 3 clients for SDBN clicks and 5 for PBM. We set the local updating time $B = 5$ with fixed global communication times $T = 10,000$. In the IID setting, at every time, each client is simulated based on a click model randomly picked from all click models instantiations detailed above and used in the non-IID setting; this provides a fair comparison between the IID and non-IID settings. We experiment on MSLR-WEB10k, Yahoo and Istella-S.

For both online and offline performance, and all datasets, our experimental results show that the difference between non-IID and IID data for Type 3 is not significant; for further details we refer the reader to the Appendix in this paper and the online appendix at \url{https://github.com/ielab/non-iid-foltr}.

\vspace{-6pt}

\subsection{Type 4: Data Quantity} 

Finally, we consider the case of \emph{data quantity skewness} (Type 4); this occurs when the number of training data varies across different clients. It is a common scenario in real-world applications. For example, in FOLTR, some clients tend to issue more queries and interact more with the searching system than others. Thus, they have more data for training than others. 
The situation represented by Type 4 may occur in combination with the other data types. In our empirical experiments, we have studied Type 4 data both on its own and combined with the document preferences skew (Type 1) and the document label distribution skew with $\#R = 1$ (Type 2).

Type 4 data is simulated by assigning different numbers of queries ($Q$) to each client during the same local updating period, thus leading to different local updating times for each client. The number of queries varies in \{1, 3, 5, 7, 9\} and we simulate $|C|=5$ clients in total with fixed global communication times $T = 10,000$. Experiments are carried out on MSLR-WEB10k, Yahoo and Istella-S.

When mixing other non-IID types with Type 4, we follow the same experimental settings of previous non-IID types, and we also assign different numbers of queries to each client (from \{1, 3, 5, 7, 9\}) during the same local updating period. Instead, in the IID simulation, each client has 5 iterations of searching for different queries. For both IID and non-IID, we use SDBN click models for click simulation and train a linear ranker using FPDGD on MSLR-WEB10k for Type 1 with $\#R = 1$, and the dataset from \citet{zhuang2021how} for Type 2.

Empirical results\footnote{In the Appendix in this paper and in the online appendix at \url{https://github.com/ielab/non-iid-foltr}.} show that if data is distributed in a non-IID manner across clients according to Type 4, the effectiveness of FPDGD is not impacted. We stress that this result may be specific to FPDGD because it uses the FedAvg paradigm and does not generalise to other FOLTR methods.

%% file: sections/conclusions.tex
\section{Outlook and Discussion}
In this paper, we provide a new perspective on the problem of data distribution across clients for federated online learning to rank. Next, we summarise the our key findings and draw directions for future research.

\textbf{Impact of non-IID data. }We found that the presence of non-IID characteristics in the distribution of document preferences (Type 1) and specific cases of document labels (Type 2) have severe effects on the effectiveness of FPDGD. Conversely, if data is distributed across clients in a non-IID manner concerning click preferences (Type 3) or data quantity (Type 4), no significant effects on the quality of FPDGD are observed. These findings contribute an understanding of under which data distributions it is safe to use FOLTR and when it is not. We believe this paper will encourage researchers to include non-IID data settings when evaluating new FOLTR methods.

\textbf{Calling for FOLTR methods to address non-IID issues. }Our paper charts directions to direct future work on non-IID data in FOLTR concerning the creation of techniques that provide remedies to Type 1 and 2, while deeming solutions for Type 3 and 4 data less critical. Importantly, we show that existing solutions employed in general federated learning to mitigate the non-IID data problem do not apply to the FOLTR setting, despite some of these non-IID cases (and especially Type 1) being likely to occur across many FOLTR systems. Thus, researching how to address non-IID data in FOLTR is a worthwhile area of investigation.

\textbf{Privacy should be a high priority when dealing with non-IID data. }Our analysis found that only the data-sharing technique could address to significant extents Type 2 non-IID data. However, this and similar methods, although performing well, require the prior knowledge about the users' local data distributions -- and thus require users to share private data, largely defeating the purpose of federated learning. 
We note that recent work has considered the sharing of synthetic, rather than real, data~\cite{wang2021non}. In such a setting, real data would be used by each client to generate synthetic data, and the synthetic data only would be shared in the federation. However, we could not find evidence of the loss in effectiveness associated with the use of synthetic rather than real data in the data-sharing scheme. Furthermore, it is unclear what the privacy guarantees are in such a synthetic data sharing scheme. Specifically, we wonder whether the use of synthetic data could jeopardise privacy as this synthetic data is generated from the real data: thus analysis of the synthetic data may reveal key aspects of and information contained in the real data.
Thus, how to guarantee user’s privacy needs when designing effective FOLTR algorithms on non-IID data is still an open question.

\textbf{Real-world datasets and benchmarks for FOLTR with non-IID data are need. }The experiments put forward in this perspective paper to substantiate our views on the non-IID data problem in FOLTR are based on simulations. While simulations are prevalent in information retrieval and especially in its evaluation~\cite{cooper1973simulation,azzopardi2016simulation,maxwell2016simulating,zhang2017information,balog2021report}, a key aspect we had to simulate was the nature of the non-IID data, including their distributions. On one hand, this allows us to carefully control the experiments; on the other it limits the generalisability of the findings to real non-IID data that may occur in FOLTR settings. We therefore want to conclude with a call for action for information retrieval practitioners in this area: there is the pressing need for FOLTR benchmark datasets that provide standard simulations on real-world non-IID scenarios as well as standard hyper-parameter settings so that future FOLTR algorithms can be fairly studied.

\section{Conclusion}

The goal of FOLTR is to learn an effective ranker in a federated (without the need for searchable and interaction data to reside on a central server) and online (by exploiting users clicks on SERPs as they occur) manner. In such a FOLTR setup, user data and interactions reside with the user's client, and not in a central server. Clients then do not need to share such data. Instead, they only share ranker updates with a central server whose responsibility is to collect such updates from the clients and aggregate them into a global model. The global model is then pushed back to the clients in an iterative manner as search interactions occur.

Despite federated learning receiving substantial attention, research in FOLTR is still in its early stages, with only two methods available at the time of writing~\cite{kharitonov2019federated,wang2021effective}. Importantly, studies that have proposed FOLTR methods have ignored an important issue that has been shown to affect the performance of federated learning systems~\cite{zhu2021federated}: that of the data not being distributed across the federated clients in an identical and independent manner (non-IID data). This paper provides the first analysis of the impact of non-IID data on FOLTR and it charts directions for future research. Our findings and observations may be valid also in other contexts that consider to create a ranker from interaction data in a federated manner, e.g., in federated counterfactual leaning to rank~\cite{li2021federated}.

We make code, experimental details and results available at \url{https://github.com/ielab/non-iid-foltr}.